\documentclass[aps,orb,twocolumn,amssymb,showpacs,pre]{revtex4-1}
\usepackage{graphicx}
\usepackage{color}
\usepackage{amstext}
\begin{document}
\def\a{{\alpha }}
\def\g{{\gamma }}
\def\b{{\beta }}
\def\z{{\zeta }}
\def\zab{{\zeta_{ab}}}
\def\be{\begin{equation}}
\def\ee#1{\label{#1}\end{equation}}
\def\d{\textsf{d} }
\def\c{\textsf{c} }
\def\e{\textsf{e} }
\def\be{\textsf{b} }
\def\s{\textsf{s} }
\def\x{\textsf{x} }
 \def\bx{\mathbf{x} }
 \def\bp{\mathbf{p} }
 \def\p{\textsf{p} }
 \def\I{\textsf{I} }
  \def\n{\textsf{n} }
 \def\pp{\textsf{P} }
 \def\q{\textsf{q} }
\def\k{\textsf{k} }
 \def\J{\textsf{J} }
\def\no{\nonumber}
\def\lb{\label}
\def\h{\textsf{h} }
\def\x{\textsf{x} }
\def\D{\textsf{D}}
\def\lb{\label}
\def\Ki{{\rm Ki}}
\def\lab{{\langle\alpha\beta\rangle}}
\newcommand{\ben}{\begin{eqnarray}}
\newcommand{\een}{\end{eqnarray}}

\title{Mixtures of relativistic gases in gravitational fields:\\ combined Chapman-Enskog and Grad method and the Onsager relations}
\author{Valdemar Moratto}\email{moratto.valdemar@gmail.com}
\affiliation{Departamento de F\'{\i}sica, Universidade Federal do
Paran\'a,  81531-980 Curitiba, Brazil}
\author{Gilberto M. Kremer}\email{kremer@fisica.ufpr.br}
\affiliation{Departamento de F\'{\i}sica, Universidade Federal do
Paran\'a,  81531-980 Curitiba, Brazil}

\begin{abstract}
In this work we study a $r$-species mixture of gases within the relativistic
kinetic theory point of view. We use the relativistic covariant
Boltzmann equation and incorporate the Schwarzschild metric. The
method of solution of the Boltzmann equation is a combination of the Chapman-Enskog and Grad
representations. The thermodynamic four-fluxes are expressed as functions of the thermodynamic forces so that the generalized expressions for the Navier-Stokes, Fick and Fourier
laws are obtained.  The constitutive equations for the diffusion  and heat four-fluxes of the mixture are functions of  thermal and  diffusion generalized forces which depend on the acceleration and the gravitational potential gradient. While this dependence is of relativistic nature for the thermal force, this is not the case for the diffusion forces. We show also that the matrix of diffusion coefficients is symmetric, implying that the thermal-diffusion equals the diffusion-thermal effect, proving the Onsager reciprocity relations. The entropy four-flow of the mixture is also expressed in terms of the thermal and diffusion generalized forces, so that its dependence on the acceleration and gravitational potential gradient is also determined.
\end{abstract}
 \maketitle

\section{Introduction}

The relativistic kinetic theory of gases is a subject that began in
1911 when J\"uttner  \cite{JUTTNER} proposed a relativistic version
of the velocity distribution function which corresponds to the Maxwellian
distribution function in the non-relativistic limiting case. Later, several studies have been made, but
for brevity's sake we mention the books \cite{GLW,CK} where several applications of the relativistic kinetic theory
of gases are discussed.

This work represents a continuation of the study of the properties
of relativistic gases using the Boltzmann equation in gravitational
fields, this sub-area has not yet been studied in depth. Here we quote some works \cite{AL,MS,KS,KD,WK} on this topic which have been recently published.

The method used in this paper to solve the covariant Boltzmann
equation is a combination of the Chapman-Enskog and Grad methods \cite{CEG,K1}.
It consists essentially in doing an expansion of the distribution
function for each species which is the solution of the Boltzmann equation up to first correction. Such a procedure is like in the Chapman-Enskog method. Then we impose that such an expansion
must be compatible with the solution of the Boltzmann equation given
by the method of Grad \cite{Grad}. In order to keep the linear regime we truncate the Grad distribution
function for each species  up to linear terms of the non-equilibrium pressure, pressure deviator tensor, diffusion and heat four-fluxes.
 Therefore, we obtain a linearized Boltzmann
equation that is written in terms of the local thermodynamic variables
and fluxes: diffusion, heat,  non-equilibrium pressure and pressure deviator tensor.
 The next step is to get
from that linearized Boltzmann equation a set of linear algebraic
system for the fluxes. We generate  one equation for each thermodynamic
flux through the multiplication of a dynamical function of
the particles by the linearized Boltzmann equation and then the integration
over the momentum space. Hence, we find the
constitutive equations for the fluxes in terms of gradients of  the local
thermodynamic variables and of a gravitational potential
that arises from the Schwarzschild metric. The laws of Navier-Stokes for the non-equilibrium pressure and pressure deviator tensor are obtained as well as the generalized Fourier and Fick laws for the heat and diffusion four-fluxes.

It will be shown that there
appears a generalized diffusion force that has,  not only dependence on the concentration and pressure gradients but also on a contribution of the four-acceleration and the gravitational potential gradient. The contributions of four-acceleration and potential gradient also appear as combined forces for the heat flux and they were analyzed separately by Eckart \cite{Eck} and Tolman \cite{To1,To2}. In the case of Eckart, for a relativistic gas in equilibrium and in the absence of gravitational fields, the temperature gradient is counterbalanced by an acceleration. On the other hand, in the case of Tolman for a relativistic gas in equilibrium and in the absence of an acceleration, the temperature gradient is counterbalanced by a gravitational potential gradient.

In order to show that the Onsager reciprocity relations hold  we manipulate the constitutive equations for the heat and diffusion fluxes. The demonstration is general in the sense that the interaction of the particles
are supposed to maintain the microscopic reversibility principle.

The structure of this paper is as follows. We define
the problem in section II and establish the Boltzmann equation and the definitions
for both, the thermodynamic variables and fluxes. In section III, we
use a method of solution of the Boltzmann equation that is a combination
of the Grad and Chapman-Enskog ones, the solution is truncated up
to first order so we obtain linear expressions. Such a process will
lead us to an algebraic system of equations for the thermodynamic
fluxes that when it is properly solved, expresses the thermodynamic coefficients
for an arbitrary inter-molecular interaction. In section IV, we show that
the Onsager reciprocity relations hold for an arbitrary inter-molecular
interaction. Furthermore,  we show that the laws of Fourier and Fick are expressed in terms of generalized thermal and diffusion forces in the presence of gravitational fields.  To give a more general representation, we show in section V that the entropy four-flow of the mixture is a function of the generalized thermal and diffusion forces. Section VI is devoted to the calculation of the constitutive equation for a relativistic Newtonian fluid, i.e., the  Navier-Stokes law.  Ultimately, in section VII, we discuss the obtained results.

\section{Background}

In this section we will define the problem of a $r$-species non-reacting
mixture in a Riemannian space with metric tensor $g^{\mu\nu}$. The
particles are supposed not have internal degrees of freedom. Each
of these particles of the constituent $a=1,...,r$ have mass $m_{a}$
and are characterized by the space-time coordinates $x^{\mu}=\left(ct,\bf x\right)$
and the momentum $p_{a}^{\mu}=\left(p_{a}^{0},\textbf{p}_{a}\right)$.
The mass-shell condition, i.e. $g_{\mu\nu}p_{a}^{\mu}p_{a}^{\nu}=m_{a}^{2}c^{2}$
implies the following relationships for the contravariant and covariant
temporal components,
\ben
\left\{
  \begin{array}{ll}
    p_{a}^{0}=(p_{a0}-g_{0i}p_{a}^{i})/{g_{00}}, \\
    p_{a0}=\sqrt{g_{00}m_{a}^{2}c^{2}+\left(g_{0i}g_{0j}-g_{00}g_{ij}\right)p_{a}^{i}p_{a}^{j}},
  \end{array}
\right.
\label{eq:1}
\een
respectively. The analysis is developed within the tenets of the
general relativity, we adopt the Schwarzschild metric $g^{\mu\nu}$
in which the line element reads \cite{Misner}:
\ben\no
ds^{2}=\left(1-\frac{2GM}{c^{2}R}\right)\left(dx^{0}\right)^{2}
-\frac{1}{\left(1-\frac{2GM}{c^{2}R}\right)}dR^{2}
\\
-R^{2}\left(d\theta^{2}+\sin^{2}\theta d\psi^{2}\right),\label{eq:3}
\een
 in terms of the spherical coordinates $\left\{ R,\theta,\psi,ct=x^{0}\right\} $.
Above, $M$ is the total mass of the spherical source and $G$ is the
gravitational constant. Here we shall use the isotropic  Schwarzschild metric, which reads
\ben\label{2}
 ds^2=g_0(r)\left(dx^0\right)^2-g_1(r)\delta_{ij}dx^idx^j,\\
g_0(r)=\frac{\left(1-\frac{GM}{2c^2r}\right)^2}{\left(1+\frac{GM}{2c^2r}\right)^2},\qquad
g_1(r)=\left(1+\frac{GM}{2c^2r}\right)^4.
\een

Along the calculation we will use a relativistic parameter
$\zeta_{a}=\frac{m_{a}c^{2}}{kT},$
 where $c$ is the speed of light, $k$ the Boltzmann constant and
$T$ the local temperature, assumed as an invariant. This parameter
is convenient because it tell us how relativistic is the system, for
example, $\zeta_{a}\gg1$ corresponds to a non-relativistic limit.
On the other hand, $\zeta_{a}\ll1$ belongs to an ultra-relativistic
limit.

The most fundamental equation in the kinetic theory is the the Boltzmann
equation; such an equation can be obtained with two hypothesis as
a basis. The first one is that particles collide elastically and only
collisions of pairs are taken into account. The second one implies
a description of the system with one-particle distribution function,
that is equivalent to think that collisions represent a process that
do not depend of what occurred in the past with the particles. This
last hypothesis is also known in the literature as molecular chaos
hypothesis. In our case, the Boltzmann equation reads \cite{CK}:
\ben\no
&&p_{a}^{\mu}\frac{\partial f_{a}}{\partial x^{\mu}}-\Gamma_{\mu\nu}^{i}p_{a}^{\mu}p_{a}^{\nu}\frac{\partial f_{a}}{\partial p_{a}^{i}}\\
&&=\sum_{b=1}^{r}\int(f_{a}'f_{b}'-f_{a}f_{b})F_{ba}\sigma_{ab}\, d\Omega\sqrt{-g}\frac{d^{3}p_{b}}{p_{b0}},\label{eq:5}
\een
 for the $a$-species. Here the Latin subindex denotes the species, note
that we have one equation with the same structure of (\ref{eq:5})
for each component of the mixture $a=1,...,r$. The distribution function
$f_{a}\left(x^{\mu},p_{a}^{\mu}\right)$ has a statistical meaning,
indeed the quantity $f_{a}\left(x^{\mu},p_{a}^{\mu}\right){d}^{3}x\,{d}^{3}p_{a}$
at time $t$, is the number of particles of the constituent $a$ in
the volume element between $\bf x$, ${\bf x}+{d}^{3}x$ and
$\textbf{p}_{a}$, $\textbf{p}_{a}+{d}^{3}p_{a}$. In equation
(\ref{eq:5}) also appear the Christoffel symbols $\Gamma_{\mu\nu}^{i}$ and
the invariant flux $F_{ba}=\sqrt{(p_{a}^{\mu}p_{b\mu})^{2}-m_{a}^{2}m_{b}^{2}c^{4}}$, which
plays the role of the relative velocity of the non-relativistic Boltzmann
equation. We have also the invariant differential elastic cross-section
$\sigma_{ab}d\Omega$ for collisions of species $a$ and $b$, where
$d\Omega$ is the corresponding solid angle element. Integrals are
made with the invariant differential element
$\sqrt{-g}\frac{d^{3}p_{b}}{p_{b0}},$
 being $\sqrt{-g}=\text{det}\left[g^{\mu\nu}\right]$. In equation
(\ref{eq:5}), quantities denoted with a prime are evaluated with
the momentum of the particles after a binary collision occurs, i.e.,
$f'_{a}\equiv f({\bf x},{\bf p}'_{a},t)$
and so on. The binary collision is characterized by the energy-momentum conservation law
$p_a^\mu+p_b^\mu=p'^\mu_a+p'^\mu_b$.

Without solving Boltzmann's equation we can obtain two important results.
The first one, arises from the H-theorem and the definition of the thermodynamic variables. A situation of local equilibrium means that the entropy four-flow production (see Eq. (\ref{s2a})) vanishes at equilibrium. The solution of the collisional term of the Boltzmann equation -- when it is equal to zero -- is the well-known local equilibrium distribution function, which reads
\begin{equation}
f_{a}^{(0)}=\frac{\n_{a}}{4\pi kTm_{a}^{2}c{K}_{2}\left(\zeta_{a}\right)} \exp\left(-\frac{U_{\mu}p_{a}^{\mu}}{kT}\right).
\label{eq:7}
\end{equation}
Here $\n_{a}$ is the local number of particles of species
$a$, the modified Bessel function of second kind is represented by
\begin{equation}
K_n(\zeta)=\left(\frac{\zeta}{2}\right)^n\frac{\Gamma(1/2)}{\Gamma(n+1/2)}\int_{1}^\infty e^{-\zeta y}\left(y^2-1\right)^{n-1/2}\,dy,
\end{equation}
and $U_{\mu}$ -- with $U^{\mu}U_{\mu}=c^{2}$ -- is the
hydrodynamical four-velocity. The set of local hydrodynamic variables
that describes the local equilibrium is $\left\{ \n_{1},...,\n_{r},U_{\mu},T\right\} $.
For the calculations that will be performed, it is convenient to evaluate
Eq. (\ref{eq:7}) in a co-moving frame, that is $U^{\mu}=\left(c/\sqrt{g_{0}},\textbf{0}\right)$
yielding
\begin{equation}
f_{a}^{(0)}=\frac{\n_{a}}{4\pi kTm_{a}^{2}c{K}_{2}\left(\zeta_{a}\right)}\exp\left(-\frac{c\sqrt{m_{a}^{2}c^{2}+g_{1}\vert{\bf p}_{a}\vert^{2}}}{kT}\right).
\end{equation}

The second important result that arises from the Boltzmann equation
is the obtention of the balance equations, for this purpose we proceed
as follows. We multiply   the Boltzmann equation (\ref{eq:5}) by the collisional invariants, that is, microscopic
dynamical quantities that are conserved between collisions, i.e., $\Psi_{a}+\Psi_{b}=\Psi_{a}'+\Psi_{b}'$
 and integrate the resulting equation over $\sqrt{-g}\frac{d^{3}p_{a}}{p_{a0}}$.
The collisional invariants $\Psi_{a}$ can take the value of the mass
and the energy-momentum of the colliding particles. To obtain the particle four-flow
balance equation for the $a$-species we take  $\Psi_{a}=c$ and integrate, this
process leads to the conservation law
\begin{equation}
N_{a;\mu}^{\mu}=0.\label{eq:8}
\end{equation}
Here the semicolon denotes a covariant derivative and we have defined
\begin{equation}
N_{a}^{\mu}=c\int p_{a}^{\mu}f_{a}\sqrt{-g}\frac{d^{3}p_{a}}{p_{a0}},\label{eq:9}
\end{equation}
as the  particle four-flow of species $a$. We now introduce a general decomposition of
$N_{a}^{\mu}$ in terms of the hydrodynamic four-velocity as
\begin{equation}
N_{a}^{\mu}=\n_{a}U^{\mu}+\J_{a}^{\mu},\qquad\hbox{where}\qquad \n_{a}=\frac{N_{a}^{\mu}U_{\mu}}{c^{2}}
\label{eq:10}
\end{equation}
 denotes the partial particle number density.
The quantity $\J_{a}^{\mu}$ is a space-like vector defined
as
\begin{equation}
\J_{a}^{\mu}=\Delta_{\nu}^{\mu}c\int p_{a}^{\nu}f_{a}\frac{d^{3}p_{a}}{p_{a0}},\label{eq:12}
\end{equation}
 and holds the property
$\J_{a}^{\mu}U_{\mu}=0$. Above, we have introduced the projector
\begin{equation}
\Delta^{\mu\nu}=g^{\mu\nu}-\frac{1}{c^{2}}U^{\mu}U^{\nu},\label{eq:14}
\end{equation}
 that has the property $\Delta^{\mu\nu}U_{\nu}=0.$
  Equation (\ref{eq:12}) is the corresponding diffusion four-flux of species
$a$ of the mixture and by taking the sum of (\ref{eq:10}) over all the components
we easily note that
\begin{equation}
N^\mu=\sum_{a=1}^r N_a^\mu=\n U^\mu,\qquad\n=\sum_{a=1}^{r}\n_{a},\qquad
\sum_{a=1}^{r}\J_{a}^{\mu}=0,
\label{eq:16}
\end{equation}
where   the last equation implies that there exist only $\left(r-1\right)$ partial diffusion
fluxes that are linearly independent for a mixture of $r$ constituents.

On the other hand, to obtain the balance equation for the energy-momentum
of the $a$-species defined by
\ben
T_a^{\mu\nu}=c\int p_{a}^{\mu}p_a^\nu f_{a}\sqrt{-g}\frac{d^{3}p_{a}}{p_{a0}},
\een
we multiply the Boltzmann equation (\ref{eq:5}) by the collisional invariant $\Psi_{a}=cp_{a}^{\mu}$
and  integrate the resulting equation over $\sqrt{-g}\frac{d^{3}p_{a}}{p_{a0}}$. This process yields
\ben\label{eq:17}
T_{a;\nu}^{\mu\nu}=P_a^\mu,
\een
where the production term $P_a^\mu$ is given by
\begin{equation}
P_a^\mu=\sum_{b=1}^{r}c\int(p_{a}^{\prime\mu}-p_{a}^{\mu})f_{a}f_{b}F_{ba}\sigma_{ab}
d\Omega\sqrt{-g}\frac{d^{3}p_{b}}{p_{b0}}\sqrt{-g}\frac{d^{3}p_{a}}{p_{a0}}.
\end{equation}
Note that this equation does not represent a conservation law, but
if we sum Eq. (\ref{eq:17}) over all species we obtain
\begin{equation}
T_{;\nu}^{\mu\nu}=\sum_{a=1}^r P_a^\mu=0,\label{eq:18}
\end{equation}
 that represents a conservation equation for the energy-momentum tensor of the mixture $T^{\mu\nu}=\sum_{a=1}^r T_a^{\mu\nu}$. By following the decomposition of Eckart (see
e.g. \cite{StWa,Heb,KP}), the energy-momentum tensor of the $a$-species
can be written as
\ben\no
T_a^{\mu\nu}=\frac{\n_a\e_a}{c^{2}}U^{\mu}U^{\nu}
+\frac{1}{c^{2}}U^{\mu}\left(\q_a^{\nu}+\h_a\J_{a}^{\nu}
\right)\\+\frac{1}{c^{2}}U^{\nu}\left(\q_a^{\mu}+\h_a\J_{a}^{\mu}
\right)-(\p_a+\varpi_a)
\Delta^{\mu\nu}+\p_a^{\langle\mu\nu\rangle},\label{eq:19}
\een
 where several definitions are to be made. First we can list the local
equilibrium quantities:  energy per particle $\e_{a}$, hydrostatic pressure $\p_{a}$ and the enthalpy per particle $\h_a=\e_{a}+\p_{a}/\n_{a}$. Next, the non-equilibrium quantities are: dynamical pressure $\varpi_a$, heat four-flux $\q^{\mu}_a$ and pressure deviator tensor $\p_a^{\langle\mu\nu\rangle}$. They are given in terms  of the following projections of the
energy-momentum tensor of the $a$-species:
\ben
\q_a^{\mu}+\h_a\J_{a}^{\mu}=\Delta_\sigma^\mu T_{a}^{\sigma\nu}U_{\nu},\qquad
\e_{a}=\frac{1}{\n_{a}c^{2}}U_{\mu}T_{a}^{\mu\nu}U_{\nu},\\
\p_a^{\langle\mu\nu\rangle}=\left(\Delta_{\sigma}^{\mu}\Delta_{\tau}^{\nu}
-\frac{1}{3}\Delta^{\mu\nu}\Delta_{\sigma\tau}\right)T_{a}^{\sigma\tau},\\
\qquad
\p_{a}+\varpi_a=-\frac13\Delta_{\mu\nu}T_{a}^{\mu\nu}.
\een
The corresponding quantities for the mixture are:
\ben\lb{sum}
\e=\sum_{a=1}^{r}\frac{\n_{a}}{\n}\e_{a},\qquad \p=\sum_{a=1}^{r}\p_{a},\qquad
\varpi=\sum_{a=1}^{r}\varpi_{a},\\\lb{sum1}
\h=\sum_{a=1}^{r}\frac{\n_{a}}{\n}\h_{a},\qquad
 \p^{\langle\mu\nu\rangle}=\sum_{a=1}^r\p_a^{\langle\mu\nu\rangle},\\\lb{sum2}
 \q^{\mu}=\sum_{a=1}^r \left(\q_a^{\mu}+\h_a\J_{a}^{\mu}\right),
\een
so that the energy-momentum tensor of the mixture is written as
\ben\no
T^{\mu\nu}=\frac{\n\e}{c^{2}}U^{\mu}U^{\nu}+\frac{1}{c^{2}}\left(U^{\mu}\q^{\nu}+U^{\nu}\q^{\mu}
\right)
\\
-(\p+\varpi)\Delta^{\mu\nu}+\p^{\langle\mu\nu\rangle}.
\een
Note that the heat four-flux $\q^{\mu}$ Eq. (\ref{sum2}) of the mixture has two contributions,
this is in accordance with the Linear Irreversible
Thermodynamics \cite{GM}, one term is related with the partial
heat flux and another with the transport of energy driven by diffusion.

Another quantity which is important in the analysis of mixtures of relativistic gases is the entropy four-flow of the mixture $S^\mu$, defined by
\ben\lb{s1}
S^\mu=-kc\sum_{a=1}^r\int p_a^\mu f_a\ln (\be_af_a) \sqrt{-g}\frac{d^{3}p_{a}}{p_{a0}},
\een
where $\be_a$ is a constant which has inverse units of $f_a$.
Its balance equation is obtained through the multiplication of the Boltzmann equation (\ref{eq:5}) by $-kc\ln (\be_af_a)$, the subsequent integration over  $\sqrt{-g}\frac{d^{3}p_{a}}{p_{a0}}$ and the sum over all species, yielding
\ben\lb{s2}
S^\mu_{;\mu}=\sigma\geq0,\\\no
\sigma=\frac{ck}4\sum_{a=1}^r\sum_{b=1}^r\int f_af_b\ln \frac{f'_af'_b}{f_af_b}\left(\frac{f'_af'_b}{f_af_b}-1\right)\\\lb{s2a}
\times F_{ba}\sigma_{ab}d\Omega\sqrt{-g}\frac{d^{3}p_{b}}{p_{b0}}\sqrt{-g}\frac{d^{3}p_{a}}{p_{a0}}.
\een
The quantity $\sigma$ is the entropy four-flow production of the mixture, which is always positive semi-definite, thanks to the relationship $ (x-1)\ln x\geq0$ valid $\forall x>0$. The entropy four-flow of the mixture is decomposed according to:
\ben\lb{s3}
S^\mu=\n\s U^\mu+\Phi^\mu, \quad \s=\frac{1}{c^2\n}S^\mu U_\mu,\quad \Phi^\mu=\Delta^\mu_\nu S^\nu,
\een
where the quantity $\s$ is identified as the entropy per particle of the mixture and $\Phi^\mu$ its entropy flux. The entropy per particle of species $a$ is given by
\ben\lb{s4}
\s_a=-\frac{k U_\mu}{c\n_a}\int p_a^\mu f_a\ln (\be_af_a)\sqrt{-g}\frac{d^{3}p_{a}}{p_{a0}},
\een
so that we have $\n\s=\sum_{a=1}^r\n_a\s_a$.

In the kinetic theory of relativistic gases there exist two decompositions that are often used: the Eckart  and the Landau-Lifshitz (see e.g. \cite{GLW,CK}). The difference between the decompositions is that the heat flux  appears in the particle four-flow  but not in the energy-momentum tensor in the Landau-Lifshitz decomposition, contrary to the Eckart one. One can take both decompositions for the determination of the constitutive equations and the results are the same. However, there are situations where one should apply only one of the decompositions, which is in the case of using BGK models of the Boltzmann collision operator. The  model equations of the Boltzmann equation normally considered in the relativistic kinetic theory are due to Marle and Anderson and Witting  (see e.g. \cite{GLW,CK}). For the Marle model one should take the Eckart decomposition, while for the Anderson and Witting model  the Landau-Lifshitz decomposition should be used.

The main problem in the kinetic theory is to find a solution of the
Boltzmann equation (\ref{eq:5}), because as we have seen, all the
above definitions can be evaluated by integrating functions that involve
$f_{a}\left(x^{\mu},p_{a}^{\mu}\right)$. The equilibrium quantities can be evaluated with the
local equilibrium distribution function  (\ref{eq:7}) and read:
\ben\lb{s5}
\e_a=m_ac^2\left(G_a-\frac1{\z_a}\right),\\
\p_a=\n_a kT,\qquad \h_a=m_ac^2G_a,\\
\s_a=k\left\{\ln\left[\frac{4\pi m_a^2ckTK_2(\z_a)}{\n_a\be_a}\right]+\z_aG_a-1\right\}.
\een
The chemical potential of species $a$ is introduced through
the Gibbs function per particle, namely, $\mu_{a}=\e_{a}-T\s_{a}+\p_{a}/\n_{a}$ and by taking into account the above expressions we get
\ben\lb{s6}
\mu_a=kT\ln\frac{e\n_a\be_a}{4\pi m_a^2ckTK_2(\z_a)}.
\een

In next sections, we will
use a method that allow to obtain expressions for the diffusion fluxes $\J_{a}^{\mu}$,
heat flux $\q^{\mu}$, non-equilibrium pressure $\varpi$, pressure deviator tensor $\p^{\langle\mu\nu\rangle}$ and entropy flux $\Phi^\mu$.
Furthermore, we will show the dependence of $\J_{a}^{\mu}$,
and  $\q^{\mu}$ in terms of the gravitational potential
and demonstrate the validity of the Onsager reciprocity relations.

\section{Combined Chapman-Enskog and Grad method}

In this section we will use a method to extract thermodynamic information from the Boltzmann equation \cite{CEG,K1}
that combines the features of the Chapman-Enskog \cite{CC} and Grad's moments one \cite{Grad}. This method has mainly two advantages, the first one is that we do not need a solution of the integro-differential Boltzmann equation as in the Chapman-Enskog method. The second is that we do not need the field equations for the moments as in the Grad method.

First we are going to describe how does the moment Grad method is
constructed. The central idea is to expand $f_{a}\left(x^{\mu},p_{a}^{\mu}\right)$
around the local equilibrium distribution function in a series of an ortho-normal
set. In this case, we have $13r+1$ unknown variables (fields) that
are described with the quantities $\left\{ \n_{a},U^{\mu},\J_{a}^{\mu},T,\varpi_{a},\q_{a}^{\mu},\p_{a}^{\left\langle \mu\nu\right\rangle }\right\} $
(see Ref. \cite{Grad}). Such an expansion reads
\begin{equation}
f_{a}=f_{a}^{(0)}\left[1+\mathcal{A}_{a}^{\mu}p_{a\mu}+\mathcal{A}_{a}^{\mu\nu}p_{a\mu}p_{a\nu}\right],\label{eq:26}
\end{equation}
where $f_{a}^{(0)}$ is the local equilibrium distribution function (J\"uttner
distribution) described by Eq. (\ref{eq:7}). In Eq. (\ref{eq:26})
the unknown tensorial coefficients $\left\{ \mathcal{A}_{a}^{\mu},\mathcal{A}_{a}^{\mu\nu}\right\} $
are calculated by solving an algebraic system constructed with the help of the definitions of
the particle four-flow $N_a^\mu$ and the energy-momentum tensor $T_a^{\mu\nu}$. The details of such calculation
are long and unnecessary to do them here, they can be consulted in Refs.
\cite{CK,KM}.
As a result of such development the distribution function $f_{a}$
will depend on linear terms of the thermodynamic
fluxes, namely,
\begin{widetext}
\ben\no
f_{a}=f_{a}^{(0)}\Bigg\{1-\frac{\J_{a\mu}}{\p_{a}}p_a^\mu
+\frac{\q_{a\mu}}{T\p_{a}}\frac{p_{a}^{\mu}}{\c_{p}^{a}}\left[\z_aG_{a}-\frac{U_{\nu}p_{a}^{\nu}}{kT}\right]+\frac{\p_{a\langle\mu\nu\rangle}}{2\p_a} \frac{\z_a}{m_a\h_a}p_a^\mu p_a^\nu
\\
+\frac{\varpi_a}{\p_a}\frac{\partial\ln\z_a}{\partial \ln\c_v^a}\Bigg[\frac{U_\mu U_\nu  p_a^\mu p_a^\nu}{k^2T^2}
-\frac{3(\c_p^a+\h_a/T)}{\c_v^a}\frac{U_\mu p_a^\mu}{kT} - \frac{\c_v^a\z_a^2+3(\c_p^a-\h_a^2/kT^2)}{\c_v^a}\Bigg]
\Bigg\}.\label{eq:27}
\een
\end{widetext}
Here we have introduced the abbreviation $G_{a}={K}_{3}\left(\zeta_{a}\right)/{K}_{2}
\left(\zeta_{a}\right)$ and the partial specific heats per particle $\c_{v}^{a}=k\left(\zeta_{a}^2+5G_{a}\z_a-G_{a}^{2}
\zeta_{a}^2-1\right)$ and $\c_p^a=\c_v^a+k$ at constant volume and pressure, respectively.

Then, following the combined Chapman-Enskog-Grad method \cite{CEG},
the expansion (\ref{eq:27}) must be compatible with the truncated
Chapman-Enskog series up to first order, that is
$f_{a}=f_{a}^{(0)}\left(1+\phi_{a}\right)$
 where $\phi_{a}$ is the first correction to the distribution function $f_{a}^{(0)}$.

Now we can proceed to linearize the Boltzmann equation as follows,
we substitute  $f_{a}=f_{a}^{(0)}\left(1+\phi_{a}\right)$ in the left hand side of the Boltzmann
equation (\ref{eq:5}) and keep the linear terms. This process is
technically the same of that developed in the Chapman-Enskog method. We also use
the so-called functional hypothesis, namely, $f_{a}=f_{a}\left(x^{\mu},p_{a}^{\mu}|\n_{a},U^{\mu},T\right)$,
leading to
\begin{widetext}
\begin{eqnarray}
p_{a}^{\mu}\frac{\partial f_{a}^{(0)}}{\partial x^{\mu}}-\Gamma_{\mu\nu}^{i}p_{a}^{\mu}p_{a}^{\mu}\frac{\partial f_{a}^{(0)}}{\partial p^{\mu}} & = & f_{a}^{(0)}\left\{ \frac{p_{a}^{\mu}}{\n_{a}}\frac{\partial\n_{a}}{\partial x^{\mu}}+\frac{p_{a}^{\mu}}{T}\left[1-\zeta_{a}G_{a}
+\frac{p_{a}^{\lambda}U_{\lambda}}{kT}\right]\frac{\partial T}{\partial x^{\mu}}-\frac{1}{kT}p_{a}^{\mu}p^i_{a}\frac{\partial U_{i}}{\partial x^{\mu}}\right. \nonumber \\
 &  & \left. -\frac{c^{2}}{2kT}\frac{p_{a}^{k}p_{a}^{i}p_{a}^{j}\delta_{ij}\delta_{kl}}
 {U^{\tau}p_{a\tau}}\frac{dg_{1}}{dr}\frac{x^{l}}{r}+\frac{c^{2}}{kT}g_{1}\Gamma_{\mu\nu}^{i}
 \frac{p_{a}^{\mu}p_{a}^{\nu}p_{a}^{j}\delta_{ij}}{U^{\tau}p_{a\tau}}\right\} .\label{eq:31-1}
\end{eqnarray}
\end{widetext}

On the other hand, we substitute the Grad function Eq. (\ref{eq:27})
in the collisional term (right hand side) of the Boltzmann equation
(\ref{eq:5}), and keep only the linear terms. This process yields
\begin{widetext}
\ben
\sum_{b=1}^{r}\int(f_{a}'f_{b}'-f_{a}f_{b})F_{ba}\sigma_{ab}\, d\Omega\sqrt{-g}\frac{d^{3}p_{b}}{p_{b0}} = -\sum_{b=1}^{r}\Bigg\{ \mathcal{I}_{ab}\left[p_{b}^{\mu}\right]\frac{\J_{b\mu}}{\p_{b}}
+\mathcal{I}_{ab}\left[p_{a}^{\mu}\right]\frac{\J_{a\mu}}{\p_{a}}
\nonumber \\\no
  -\mathcal{I}_{ab}\left[\frac{p_{b}^{\mu}}{\c_{p}^{b}}\left(\zeta_{b}G_{b}
-\frac{U_{\nu}p_{b}^{\nu}}{kT}\right)\right]\frac{\q_{b\mu}}{T\p_{b}}
 -\mathcal{I}_{ab}\left[\frac{p_{a}^{\mu}}{\c_{p}^{a}}\left(\zeta_{a}G_{a}
 -\frac{U_{\nu}p_{a}^{\nu}}{kT}\right)\right]
 \frac{\q_{a\mu}}{T\p_{a}}-\mathcal{I}_{ab}\left[\frac{\z_b}{m_b\h_b}p_b^\mu p_b^\nu\right]
 \frac{\p_{b\langle\mu\nu\rangle}}{2\p_b}
 \\\no
 -\mathcal{I}_{ab}\left[\frac{\z_a}{m_a\h_a}p_a^\mu p_a^\nu\right]
 \frac{\p_{a\langle\mu\nu\rangle}}{2\p_a}-\mathcal{I}_{ab}\left[\frac{\partial\ln\z_b}{\partial \ln\c_v^b}\left(\frac{U_\mu U_\nu  p_b^\mu p_b^\nu}{k^2T^2}
-\frac{3(\c_p^b+\h_b/T)}{\c_v^b}\frac{U_\mu p_b^\mu}{kT}\right)\right]
 \frac{\varpi_b}{\p_b}
 \\
-\mathcal{I}_{ab}\left[\frac{\partial\ln\z_a}{\partial \ln\c_v^a}\left(\frac{U_\mu U_\nu  p_a^\mu p_a^\nu}{k^2T^2}
-\frac{3(\c_p^a+\h_a/T)}{\c_v^a}\frac{U_\mu p_a^\mu}{kT}\right)\right]
 \frac{\varpi_a}{\p_a}
 \Bigg\} .\label{eq:32-1}
\een
\end{widetext}
 Here we have introduced the collision operators
\begin{equation}
\mathcal{I}_{ab}\left[\varphi_{a}\right]=\int f_{a}^{(0)}f_{b}^{(0)}\left(\varphi_{a}^{\prime}
-\varphi_{a}\right)F_{ab}\sigma_{ab}d\Omega\sqrt{-g}\frac{d^{3}p_{b}}{p_{b0}},\label{eq:31}
\end{equation}
 for any function that depends on the momentum four-vector $\varphi_{a}(p_a^{\mu})$. Note that
Eq. (\ref{eq:31}) imply that we can write for an arbitrary function $\psi_b(p_b^\mu)$
\begin{equation}
\int\psi_{b}\mathcal{I}_{ab}\left[\varphi_{a}\right]\sqrt{-g}\frac{d^{3}p_{a}}{p_{a0}}
=\int\varphi_{a}\mathcal{I}_{ab}\left[\psi_{b}\right]\sqrt{-g}\frac{d^{3}p_{a}}{p_{a0}},\label{eq:32}
\end{equation}
 thanks to the symmetry properties of the collision operator.

By collecting the above information the linearized Boltzmann equation in the  combined Chapman-Enskog-Grad method
becomes
\begin{widetext}
\ben
f_{a}^{(0)}\Bigg\{ \frac{p_{a}^{\mu}}{\n_{a}}\frac{\partial\n_{a}}{\partial x^{\mu}}+\frac{p_{a}^{\mu}}{T}\left[1-\zeta_{a}G_{a}
+\frac{p_{a}^{\lambda}U_{\lambda}}{kT}\right]\frac{\partial T}{\partial x^{\mu}}-\frac{1}{kT}p_{a}^{\mu}p^i_{a}\frac{\partial U_{i}}{\partial x^{\mu}}-\frac{c^{2}}{2kT}\frac{p_{a}^{k}p_{a}^{i}p_{a}^{j}\delta_{ij}\delta_{kl}}
 {U^{\tau}p_{a\tau}}\frac{dg_{1}}{dr}\frac{x^{l}}{r}
 \nonumber
  +\frac{c^{2}}{kT}g_{1}\Gamma_{\mu\nu}^{i}
 \frac{p_{a}^{\mu}p_{a}^{\nu}p_{a}^{j}\delta_{ij}}{U^{\tau}p_{a\tau}}\Bigg\}
 \\\no= -\sum_{b=1}^{r}\Bigg\{ \mathcal{I}_{ab}\left[p_{b}^{\mu}\right]\frac{\J_{b\mu}}{\p_{b}}
+\mathcal{I}_{ab}\left[p_{a}^{\mu}\right]\frac{\J_{a\mu}}{\p_{a}}-\mathcal{I}_{ab}
\left[\frac{p_{b}^{\mu}}{\c_{p}^{b}}\left(\zeta_{b}G_{b}
-\frac{U_{\nu}p_{b}^{\nu}}{kT}\right)\right]\frac{\q_{b\mu}}{T\p_{b}}
 -\mathcal{I}_{ab}\left[\frac{p_{a}^{\mu}}{\c_{p}^{a}}\left(\zeta_{a}G_{a}
 -\frac{U_{\nu}p_{a}^{\nu}}{kT}\right)\right]
 \frac{\q_{a\mu}}{T\p_{a}}
\nonumber \\\no
  -\mathcal{I}_{ab}\left[\frac{\z_b}{m_b\h_b}p_b^\mu p_b^\nu\right]
 \frac{\p_{b\langle\mu\nu\rangle}}{2\p_b}
  -\mathcal{I}_{ab}\left[\frac{\z_a}{m_a\h_a}p_a^\mu p_a^\nu\right]
 \frac{\p_{a\langle\mu\nu\rangle}}{2\p_a}-\mathcal{I}_{ab}\left[\frac{\partial\ln\z_b}{\partial \ln\c_v^b}\left(\frac{U_\mu U_\nu  p_b^\mu p_b^\nu}{k^2T^2}
-\frac{3(\c_p^b+\h_b/T)}{\c_v^b}\frac{U_\mu p_b^\mu}{kT}\right)\right]
 \frac{\varpi_b}{\p_b}
 \\
-\mathcal{I}_{ab}\left[\frac{\partial\ln\z_a}{\partial \ln\c_v^a}\left(\frac{U_\mu U_\nu  p_a^\mu p_a^\nu}{k^2T^2}
-\frac{3(\c_p^a+\h_a/T)}{\c_v^a}\frac{U_\mu p_a^\mu}{kT}\right)\right]
 \frac{\varpi_a}{\p_a}
 \Bigg\},\qquad\label{eq:38-1}
 \een
\end{widetext}
due to (\ref{eq:31-1}) and (\ref{eq:32-1}).

In the next sections we will use (\ref{eq:38-1}) in order to determine the constitutive equations for the diffusion fluxes $\J_{a}^{\mu}$,
heat flux $\q^{\mu}$, non-equilibrium pressure $\varpi$ and pressure deviator tensor $\p^{\langle\mu\nu\rangle}$.

\section{Fick and Fourier laws}
 Now we will obtain a system of linear equations for the determination of the  the diffusion fluxes  $\J_{a}^\mu$ and the heat flux of the mixture $\q^{\mu}$. The solution of
such a system will represent the form of the linear fluxes in terms
of the thermodynamic forces. The integral functions for the transport
coefficients and therefore the Onsager reciprocity relations will
be analyzed in the next subsection.

To obtain the first one of the looked set of equations, we multiply
Eq. (\ref{eq:38-1}) by $c\Delta_{\nu}^{\mu}p_{a}^{\nu}/\n_{a}$
and integrate over $\sqrt{-g}\frac{d^{3}p_{a}}{p_{a0}}$. The integrals used for this process can be consulted in the Appendix. The resulting equation
is
\ben\no
-\frac{1}{\n_{a}}\nabla^\mu\p_{a}+\frac{\h_{a}}{c^{2}}\Delta^{\mu i}\left[U^{\nu}\frac{\partial U_{i}}{\partial x^{\nu}}-\frac{1}{1-{\Phi^2}/4c^{4}}\frac{\partial{\Phi}}{\partial x^{i}}\right]\\
=\sum_{b=1}^{r}\left(\mathcal{A}_{ab}\J_{b}^{\mu}-\mathcal{F}_{ab}\q_{b}^{\mu}\right),\label{eq:35}
\een
 where $\nabla^\mu=\Delta^{\mu\nu}\partial_\nu$ is the gradient operator and ${\Phi}=-\frac{GM}{r}$ is the gravitational potential. In Eq. (\ref{eq:35}) we have
introduced the matrices $\mathcal{A}_{ab}$ and $\mathcal{F}_{ab}$.
We can split  $\mathcal{A}_{ab}$ for different indices $\left\{ a,b\right\} $
\begin{equation}
\mathcal{A}_{ab}=-\frac{c\Delta^{\mu\nu}}{3\n_{a}\n_{b}kT}\int p_{a\mu}\mathcal{I}_{ab}\left[p_{b\nu}\right]\sqrt{-g}\frac{d^{3}p_{a}}{p_{a0}},\quad a\neq b,\label{eq:36}
\end{equation}
 and for equal indices $\left\{ a,b=a\right\} $,
\ben\no
\mathcal{A}_{aa}=-\frac{c\Delta^{\mu\nu}}{3\n_{a}^{2}kT}\bigg[\sum_{b=1}^{r}\int p_{a\mu}\mathcal{I}_{ab}\left[p_{a\nu}\right]\\+\int p_{a\mu}\mathcal{I}_{aa}\left[p_{a\nu}\right]\bigg]\sqrt{-g}\frac{d^{3}p_{a}}{p_{a0}}.\label{eq:37}
\een
 The matrix $\mathcal{F}_{ab}$ introduced in Eq. (\ref{eq:35})
is written by doing the same splitting, for unlike indices $\left\{ a,b\right\} $
we have
\ben\no
\mathcal{F}_{ab}=-\frac{c\Delta^{\mu\nu}}{3\n_{a}\n_{b}kT^{2}}\int p_{a\mu}\mathcal{I}_{ab}\left[\frac{\zeta_{b}}{\c_{p}^{b}}
\left(G_{b}-\frac{U_{\tau}p_{b}^{\tau}}{m_{b}c^{2}}\right)p_{b\nu}\right]
\\\times\sqrt{-g}\frac{d^{3}p_{a}}{p_{a0}},\quad a\neq b,\qquad\label{eq:38}
\een
 and for like indices $\left\{ a,b=a\right\} $,
\ben
\mathcal{F}_{aa} = -\frac{c\Delta^{\mu\nu}}{3\n_{a}^{2}kT^{2}}\left\{ \sum_{b=1}^{r}\int p_{a\mu}\mathcal{I}_{ab}\left[\frac{\zeta_{a}}{\c_{p}^{a}}
\left(G_{a}-\frac{U_{\tau}p_{a}^{\tau}}{m_{a}c^{2}}\right)p_{a\nu}\right]
\right.\nonumber \\
  \left.+\int p_{a\mu}\mathcal{I}_{aa}\left[\frac{\zeta_{a}}
  {\c_{p}^{a}}\left(G_{a}-\frac{U_{\tau}p_{a}^{\tau}}
 {m_{a}c^{2}}\right)p_{a\nu}\right]\right\}\sqrt{-g}\frac{d^{3}p_{a}}{p_{a0}}.\qquad\label{eq:39}
\een

Next we look for a second equation which is independent from Eq. (\ref{eq:35}).
Hence, we multiply the linearized Boltzmann equation (\ref{eq:38-1})
by $\Delta_{\nu}^{\mu}\frac{c\zeta_{a}}{\c_{p}^{a}\n_{a}T}\left(G_{a}-\frac{U_{\sigma}p_{a}^{\sigma}}{m_{a}c^{2}}\right)p_{a}^{\nu}$
and integrate over $\sqrt{-g}\frac{d^{3}p_{a}}{p_{a0}}$, for this
long process we use also the integrals that appear in the Appendix.
The result becomes
\ben\no
\frac{1}{T}\left\{ \nabla^\mu T-\frac{T}{c^{2}}\Delta^{\mu i}\left[U^{\nu}\frac{\partial U_{i}}{\partial x^{\nu}}-\frac{1}{1-{\Phi^2}/4c^{4}}\frac{\partial{\Phi}}{\partial x^{i}}\right]\right\} \\=\sum_{b=1}^{r}\left(\mathcal{F}_{ba}\J_{b}^{\mu}-\mathcal{H}_{ab}\q_{b}^{\mu}\right),\label{eq:40}
\een
 where another matrix $\mathcal{H}_{ab}$ is defined. As
with the others operators, we split $\mathcal{H}_{ab}$ in the part
for unlike indices $\left\{ a,b\right\} $
\ben\no
\mathcal{H}_{ab}=-\frac{c\Delta^{\mu\nu}}{3\n_{a}\n_{b}kT^{3}}
\int\frac{\zeta_{a}}{\c_{p}^{a}}\left(G_{a}-\frac{U_{\sigma}p_{a}^{\sigma}}
{m_{a}c^{2}}\right)p_{a\mu}\\\times\mathcal{I}_{ab}\left[\frac{\zeta_{b}}{\c_{p}^{b}}\left(G_{b}-\frac{U_{\epsilon}p_{b}^{\epsilon}}{m_{b}c^{2}}\right)p_{b\nu}\right]\sqrt{-g}\frac{d^{3}p_{a}}{p_{a0}},\quad a\neq b,\label{eq:41}
\een
 and the corresponding for same indices
\ben\no
\mathcal{H}_{aa} = -\frac{c\Delta^{\mu\nu}}{3\n_{a}^{2}kT^{3}}\Bigg\{ \sum_{b=1}^{r}\int\frac{\zeta_{a}}{\c_{p}^{a}}\left(G_{a}-\frac{U_{\sigma}p_{a}^{\sigma}}
{m_{a}c^{2}}\right)p_{a\mu}
\\\times
\mathcal{I}_{ab}\left[\frac{\zeta_{a}}{\c_{p}^{a}}
\left(G_{a}-\frac{U_{\epsilon}p_{a}^{\epsilon}}{m_{a}c^{2}}\right)p_{a\nu}\right]
+\int\frac{\zeta_{a}}{\c_{p}^{a}}\left(G_{a}-\frac{U_{\sigma}p_{a}^{\sigma}}
 {m_{a}c^{2}}\right)p_{a\mu}
\nonumber \\
 \times\mathcal{I}_{aa}\left[\frac{\zeta_{a}}{\c_{p}^{a}}
\left(G_{a}-\frac{U_{\epsilon}p_{a}^{\epsilon}}{m_{a}c^{2}}\right)p_{a\nu}\right]
\Bigg\} \sqrt{-g}\frac{d^{3}p_{a}}{p_{a0}}.\qquad\label{eq:42}
\een

Hence, we have obtained the desired system of algebraic equations, namely (\ref{eq:35})
and (\ref{eq:40}) which are an independent set of linear equations for the determination of the diffusion $\J_{a}^{\mu}$ and heat $\q_{a}^{\mu}$ fluxes.

\subsection{Onsager reciprocity relations}

In this section we will show that the Onsager reciprocity relations hold
for the system under consideration. The idea is to verify if the matrix associated
with the diffusion coefficients is symmetric and therefore the
so-called cross effects are equal as it is described from one of the hypothesis
of the Linear Irreversible Thermodynamics \cite{GM}. One
cross effect for our system is the contribution to diffusion
due to the temperature gradient, this is often called ``Soret'' effect.
The other cross effect is the contribution to the heat flux due to
the chemical potential gradient or a concentration gradient, when
it is due to the last, it is called ``Dufour'' effect. This demonstration
is general in the sense that no interaction between the particles
is established, but of course, the microscopic reversibility principle
is called for the collisional term of the Boltzmann equation (\ref{eq:5}).

Let us now write the thermodynamic forces in order to identify
clearly the Soret and Dufour effects in terms of the temperature and chemical potential gradients.

First we define a generalized thermal force as
\begin{equation}\lb{td1}
\nabla^{\mu}\mathcal{T}=\nabla^\mu T-\frac{T}{c^{2}}\Delta^{\mu i}\left[U^{\nu}\frac{\partial U_{i}}{\partial x^{\nu}}-\frac{1}{1-{\Phi^2}/4c^{4}}\frac{\partial{\Phi}}{\partial x^{i}}\right],
\end{equation}
where the first term contains a temperature gradient  while the second  one -- whose nature is strictly relativistic due to the factor $T/c^{2}$ -- is proportional to the four-acceleration and the gravitational
potential gradient. The term due to the four-acceleration was proposed by Eckart \cite{Eck} while the one due to
the gravitational potential gradient by Tolman \cite{To1,To2}. If we think in a relativistic gas in equilibrium, we can conjecture the following two aspects: (i) in the absence of a gravitational potential gradient, the temperature gradient must be counterbalanced by an acceleration  and  (ii) in the absence of an acceleration, the temperature gradient must be counterbalanced by a gravitational potential gradient. Now, Eq. (\ref{eq:40}) can be  written in terms of the thermal force as
\ben\lb{td2}
\frac1{T}\nabla^{\mu}\mathcal{T}=
\sum_{b=1}^{r-1}\left(\mathcal{F}_{ba}-\mathcal{F}_{ra}\right)\J_{b}^{\mu}
-\sum_{b=1}^{r}\mathcal{H}_{ab}\q_{b}^{\mu}.
\een
Above we have considered the constraint $\sum_{a=1}^r\J_{a}^{\mu}=0$ which implies that there exist only $r-1$ linearly independent diffusion fluxes.

Next we recall that the chemical potential of species $a$ is defined through
the Gibbs function per particle ($\mu_{a}=\e_{a}-T\s_{a}+\p_{a}/\n_{a}$).
So that, the following important relationship holds for its gradients
\begin{equation}
\nabla^\mu\left(\frac{\mu_{a}}{T}\right)=\frac{1}{\n_{a}T}\nabla^\mu\p_{a}-\frac{\h_{a}}{T^{2}}
\nabla^\mu T.\label{eq:55}
\end{equation}
Therefore, the substitution of Eq. (\ref{eq:55}) into (\ref{eq:35}) yields
\begin{equation}
-T\nabla^\mu\left(\frac{\mu_{a}}{T}\right)-\frac{\h_{a}}{T}\nabla^{\mu}\mathcal{T} =\sum_{b=1}^{r}\left(\mathcal{A}_{ab}\J_{b}^{\mu}-\mathcal{F}_{ab}\q_{b}^{\mu}\right).\label{eq:56}
\end{equation}
Moreover, by considering that there exist $r-1$ independent diffusion
fluxes, we we can take the $r$th
component of Eq. (\ref{eq:56}) and subtract it from (\ref{eq:56})
itself, yielding
\ben\no
-T\nabla^\mu\left(\frac{\mu_{a}-\mu_{r}}{T}\right)-\frac{\h_{a}-\h_{r}}{T}\nabla^{\mu}\mathcal{T}=
\\\no
=\sum_{b=1}^{r-1}\left(\mathcal{A}_{ab}-\mathcal{A}_{rb}-\mathcal{A}_{ar}+\mathcal{A}_{rr}\right)
\J_{b}^{\mu}\\\label{eq:45}-\sum_{b=1}^{r}\left(\mathcal{F}_{ab}-\mathcal{F}_{rb}\right)\q_{b}^{\mu}.
\een

Now we can proceed to solve the system of linear equations formed
by Eqs. (\ref{td2}) and (\ref{eq:45}). First we solve Eq. (\ref{td2})
for $\q_{b}^{\mu}$, yielding
\ben\no
\q_{c}^{\mu} = \sum_{d=1}^{r}\left(\mathcal{H}^{-1}\right)_{cd}\left\{ -\frac{1}{T}\nabla^{\mu}\mathcal{T}\right\} \\+\sum_{d=1}^{r}\sum_{b=1}^{r-1}\left(\mathcal{H}^{-1}\right)_{cd}
\left(\mathcal{F}_{bd}-\mathcal{F}_{rd}\right)\J_{b}^{\mu},\label{eq:46}
\een
 where $\left(\mathcal{H}^{-1}\right)_{cd}$ is the inverse matrix
of $\mathcal{H}_{cd}$ so that $\left(\mathcal{H}^{-1}\right)_{cd}\mathcal{H}_{da}=\delta_{ca}$
is the identity matrix. Then we insert Eq. (\ref{eq:46}) into Eq.
(\ref{eq:45}) and solve for $\J_{a}^{\mu}$,
\begin{equation}
\J_{a}^{\mu}=-T\sum_{b=1}^{r-1}\mathcal{D}'_{ab}\nabla^\mu\left(\frac{\mu_{b}-\mu_{r}}{T}\right)
-\frac{\mathcal{D}_{a}}{T}\nabla^{\mu}\mathcal{T}.\label{eq:47}
\end{equation}
Here we identify the above equation as the generalized Fick law, where the coefficients $\mathcal{D}'_{ab}$ and $\mathcal{D}_{a}$ are related with the diffusion and thermal-diffusion (Soret) effects, respectively. The inverse of the diffusion matrix reads
\ben\no
\left(\mathcal{D}^{'-1}\right)_{ab}=\mathcal{A}_{ab}-\mathcal{A}_{rb}-\mathcal{A}_{ar}
+\mathcal{A}_{rr}
\\
-\sum_{c=1}^{r}\sum_{d=1}^{r}
\left(\mathcal{F}_{ac}-\mathcal{F}_{rc}\right)
\left(\mathcal{H}^{-1}\right)_{cd}\left(\mathcal{F}_{bd}-\mathcal{F}_{rd}\right),\label{eq:49}
\een
while the thermal-diffusion coefficients are given by
\begin{equation}
\mathcal{D}_{a}=\sum_{b=1}^{r-1}\mathcal{D}'_{ab}\left\{ \h_{b}-\h_{r}+\sum_{c=1}^{r}\sum_{d=1}^{r}\left(\mathcal{F}_{bc}-\mathcal{F}_{rc}\right)\left(\mathcal{H}^{-1}\right)_{cd}\right\} .\label{eq:48}
\end{equation}

Now we have to obtain the total heat flux as a function of the temperature and chemical potential gradients. For this end, we rewrite the total heat four-flux (\ref{sum2}) as
\begin{equation}
\q^{\mu}=\sum_{a=1}^{r}\q_{a}^{\mu}+\sum_{a=1}^{r-1}\left(\h_{a}-\h_{r}\right)\J_{a}^{\mu},\label{eq:50}
\end{equation}
and substitute in it the expressions found for $\q_{a}^{\mu}$ and $\J_{a}^{\mu}$, i.e. Eqs. (\ref{eq:46}) and (\ref{eq:47}). Hence it follows the Fourier law
\begin{equation}
\q^{\mu}=-\frac{\lambda'}{T}\nabla^{\mu}\mathcal{T}-T\sum_{a=1}^{r-1}\mathcal{D}'_{a}\nabla^\mu\left(\frac{\mu_{a}-\mu_{r}}{T}\right),\label{eq:51}
\end{equation}
 where we have introduced the thermal conductivity coefficient
\ben\no
\lambda'=\sum_{a=1}^{r}\sum_{b=1}^{r}\left(\mathcal{H}^{-1}\right)_{ab}
+\sum_{b=1}^{r-1}\mathcal{D}_{b}\Bigg[\h_{b}-\h_{r}
\\
+\sum_{a=1}^{r}\sum_{c=1}^{r}\left(\mathcal{H}^{-1}\right)_{ac}
\left(\mathcal{F}_{bc}-\mathcal{F}_{rc}\right)\Bigg],\label{eq:52}
\een
 and the diffusion-thermal coefficient
\begin{equation}
\mathcal{D}'_{a}=\sum_{b=1}^{r-1}\mathcal{D}'_{ba}\left[\h_{b}-\h_{r}+\sum_{c=1}^{r}\sum_{d=1}^{r}\left(\mathcal{H}^{-1}\right)_{cd}\left(\mathcal{F}_{bd}-\mathcal{F}_{rd}\right)\right].\label{eq:53}
\end{equation}

Ultimately, we make a close inspection of the matrices $\mathcal{A}_{ab}$,  $\mathcal{F}_{ab}$ and $\mathcal{H}_{ab}$ which are given as functions of the
collision operators $\mathcal{I}_{ab}$. From (\ref{eq:36}), (\ref{eq:38}) and (\ref{eq:41}), we may infer that only $\mathcal{A}_{ab}$  and $\mathcal{H}_{ab}$ are symmetric matrices, while $\mathcal{F}_{ab}$ is non-symmetric. Hence we may conclude from  (\ref{eq:49}) that the matrix related with the diffusion coefficients are symmetric, i.e., $\mathcal{D}'_{ab}=\mathcal{D}'_{ba}$.  Moreover,  for the coefficients of cross effects -- namely the Soret $\mathcal{D}_{a}$  and  Dufour $\mathcal{D}'_{a}$
  -- we note from the symmetry of
 $\mathcal{H}_{ab}$ and $\mathcal{D}_{ab}$ that (\ref{eq:48}) and (\ref{eq:53}) are equivalent, so that $\mathcal{D}_{a}=\mathcal{D}'_{a}$.
 The relationships  $\mathcal{D}'_{ab}=\mathcal{D}'_{ba}$ and $\mathcal{D}_{a}=\mathcal{D}'_{a}$ imply a demonstration of the validity of the Onsager reciprocity
relations. Note that as in another's demonstrations \cite{Valdemar},
it appears to be an ultimate macroscopic effect that can only be proved
because of the symmetries that belong to the collisional term of the
Boltzmann equation (\ref{eq:5}) given from the H-theorem i.e. microscopic
reversibility principle.

\subsection{Thermal and diffusion forces}

It is usual in the theory of fluid mixtures to express the diffusion fluxes and the heat flux of the mixture in terms of the generalized thermal and diffusion forces. The thermal force for a relativistic fluid was introduced in the last section (see (\ref{td1})). On the other hand, we follow \cite{KD} and define the generalized diffusion force of species $a$ as
\ben\no
&&\d_a^{\mu}=\nabla^\mu\x_{a}+\left(\x_{a}-1\right)\nabla^\mu\ln\p
\\\lb{td3}
&&-\frac{\n_{a}\h_{a}-\n\h}{\p c^{2}}\Delta^{\mu j}\left[U^{\tau}\frac{\partial U_{j}}{\partial x^{\tau}}-\frac{1}{1-{\Phi^2}/4c^{4}} \frac{\partial{\Phi}}{\partial x^{j}}\right],
\een
where $\x_{a}=\p_{a}/\p=\n_{a}/\n$ is the concentration of species $a$. We can identify four contributions to the generalized diffusion force: a concentration gradient, a pressure gradient, a term proportional to the four-acceleration and the gradient of the gravitational potential. Here it is important to emphasize that contrary to what happens with the thermal force, the terms with the four-acceleration and gradient of gravitational potential are not of a strictly relativistic nature. Indeed, $(\n_a\h_a-\n\h)/c^2\p=(\n_am_aG_a-\sum_{b=1}^r\n_bm_bG_b)/\p$ and $G_b\rightarrow1$ for $\z_b\gg1$. This  equation  has a very important feature because
it represents the generalization of the diffusion force originally written for the non-relativistic
case \cite{CC,HCB}. More discussions about this point can be found in \cite{KD}. Here as in the non-relativistic case, exist only $r-1$ linearly independent generalized diffusion forces due to the relationship $\sum_{a=1}^r\d_a^\mu=0$.

Now we can proceed to express the vectorial fluxes in terms of the generalized thermal and diffusive forces. To do so, we use the momentum density balance equation (see \cite{KD})
\begin{equation}
\frac{\partial\p}{\partial x^{i}}-\frac{\n\h}{c^{2}}\left[U^{\nu}\frac{\partial U_{i}}{\partial x^{\nu}}-\frac{1}{1-{\Phi^2}/4c^{4}}\frac{\partial{\Phi}}{\partial x^{i}}\right]=0,\label{eq:49-1}
\end{equation}
and the gradient of the chemical potential written as
\ben\no
-T\nabla^\mu\left(\frac{\mu_{a}-\mu_{r}}{T}\right)=
-\frac{\p}{\n_a}\left(\nabla^\mu\x_a+\x_a\nabla^\mu\ln\p\right)
\\
+\frac{\p}{\n_r}\left(\nabla^\mu\x_r+\x_r\nabla^\mu\ln\p\right)+\frac{\h_a-\h_r}{T}\nabla^\mu T.
\een
After some rearrangements, the expressions for (\ref{eq:47}) and  (\ref{eq:51}) become
\ben\lb{Jq1}
\J_a^\mu=\sum_{b=1}^{r-1}\widetilde{\mathcal{D}}_{ab}\d_b^\mu
+\frac{\widetilde{\mathcal{D}}_a}{T}\nabla^\mu\mathcal{T},\\
\lb{Jq11}\q^\mu=\frac{\widetilde\lambda}{T}\nabla^\mu\mathcal{T}
+\sum_{a=1}^{r-1}\widetilde{\mathcal{D}}'_{a}\d_a^\mu.
\een
In these representations for the generalized thermal and diffusion forces, the transport coefficients read:
\ben\lb{Jq2}
&&\widetilde{\mathcal{D}}_{ab}=-\sum_{c=1}^{r-1}\mathcal{D}'_{ac}\frac{\p}{\n_b}\left(\delta_{bc}
+\frac{\n_b}{\n_r}\right),
\\ &&\widetilde{\mathcal{D}}_a=-\sum_{b=1}^{r-1}\sum_{c=1}^{r}\sum_{d=1}^{r}\mathcal{D}'_{ab}
\left(\mathcal{F}_{bc}-\mathcal{F}_{rc}\right)\left(\mathcal{H}^{-1}\right)_{cd},
\\\no\lb{Jq3}
&&\widetilde\lambda=-\sum_{a=1}^{r}\sum_{b=1}^{r}\Bigg[\left(\mathcal{H}^{-1}\right)_{ab}
\\&&+\sum_{c=1}^{r-1}\mathcal{D}_{c}\left(\mathcal{H}^{-1}\right)_{ab}
\left(\mathcal{F}_{cb}-\mathcal{F}_{rb}\right)\Bigg],\\ &&\widetilde{\mathcal{D}}'_{a}=-\sum_{b=1}^{r-1}\mathcal{D}_a\frac{\p}{\n_a}
\left(\delta_{ab}+\frac{\n_a}{\n_r}\right).
\een

At this point it is worth pausing to make two comments. Firstly, by looking the expression for the generalized diffusion force Eq. (\ref{td3}) we note that it depends on: (i) A concentration gradient, that tends to reduce the non-homogeneity of the mixture; (ii) a pressure gradient, where heavy particles tend to diffuse to places with high pressures, e.g. in centrifuges; (iii) an acceleration, which acts on different masses and (iv) a gravitational potential gradient. Secondly, let us suppose a mixture in which the generalized thermal force vanishes, the pressure is constant and there is no acceleration. We can think also that there is no diffusive flux, implying a pseudo-equilibrium state. It is very interesting such a situation because of its physical implications, that is,  the gradient of concentration has to be counterbalanced by the gravitational potential gradient.

To complete this section, we point out that the thermal conductivity coefficient $\lambda$ in a mixture is defined as the ratio of the heat flux to the temperature gradient. This when there is no diffusion i.e., when $\J_a^\mu=0$. From (\ref{eq:46}) and (\ref{eq:50}), we have
\ben
\lambda=\sum_{a=1}^{r}\sum_{b=1}^{r}\left(\mathcal{H}^{-1}\right)_{ab}.
\een
Furthermore, in the absence of a temperature gradient the constitutive equation for the diffusion fluxes (\ref{Jq1}) are proportional only to the generalized diffusion forces and   $\widetilde{\mathcal{D}}_{ab}$ is identified as the matrix of the diffusion coefficients.

\section{Entropy flux of the mixture}

In this section we will show that the entropy four-flow for the system under consideration takes the form as predicted by the Linear Irreversible Thermodynamics. According to (\ref{s1}) and (\ref{s3})$_3$ the entropy flux of the mixture is given by
\ben\lb{en1}
\Phi^\mu =-kc\Delta^\mu_\nu\sum_{a=1}^r\int p_a^\nu f_a\ln (\be_af_a) \sqrt{-g}\frac{d^{3}p_{a}}{p_{a0}}.
\een
We substitute the Grad distribution function (\ref{eq:27}) into the above expression and linearize in the fluxes $\J_{a}^{\mu}, \q^{\mu}, \varpi,\p^{\langle\mu\nu\rangle}$. After integration the entropy flux takes the form
\ben\no
\Phi^\mu=\frac1{T}\sum_{a=1}^r\q_a^\mu+\sum_{a=1}^r\s_a\J_a^\mu
=\frac{\q^\mu}{T}-\sum_{a=1}^r\frac{\mu_a}{T}\J_a^\mu
\\\lb{en2}
=\frac{\q^\mu}{T}-\sum_{a=1}^{r-1}\frac{\mu_a-\mu_r}{T}\J_a^\mu.
\een
The first equality above shows that the entropy flux of the mixture is a sum of two terms: one refers to the sum of all partial heat fluxes divided by the temperature and the other is a sum of the transport  due to diffusion of the partial entropies per particle. The second equality is well-known from non-relativistic Linear Irreversible Thermodynamics  \cite{GM}, and is connected with the transport of the chemical potentials driven by diffusion.

We can also express the entropy flux of the mixture in terms of the thermal and diffusion generalized forces by substituting the representations (\ref{Jq1}) and (\ref{Jq11}) into (\ref{en2}), yielding
\ben\lb{en3}
\Phi^\mu=-\frac{\mathcal{L}}{T^2}\nabla^\mu\mathcal{T}-\sum_{a=1}^{r-1}\frac{\mathcal{L}_a}{T}\d_a^\mu.
\een
Here the scalar coefficients $\mathcal{L}$ and $\mathcal{L}_a$ read
\ben\lb{en4}
\mathcal{L}=\widetilde\lambda-\sum_{a=1}^{r-1}\left(\mu_a-\mu_r\right)\widetilde{\mathcal{D}}_a,\\
\mathcal{L}_a=\widetilde{\mathcal{D}}'_a-\sum_{b=1}^{r-1}\left(\mu_b-\mu_r\right)\widetilde{\mathcal{D}}_{ba}.
\een
It is clear from the definitions of the thermal (\ref{td1}) and diffusive (\ref{td3}) forces that the entropy flux of the mixture (\ref{en3}) depends on the temperature, concentration and pressure gradients as well as on the acceleration and gravitational potential gradient.

\section{Navier-Stokes law}

In this section we will calculate the constitutive equations for a relativistic Newtonian fluid, in other words the Navier-Stokes law. This law is usually separated in two equations. The first one is for the non-equilibrium pressures and it is associated with the bulk viscosity. The second one is for the pressure deviator tensor and it is associated with the shear viscosity.

Let us start with the constitutive equation for the partial non-equilibrium pressures $\varpi_a$ of species $a$, it is obtained as follows. We multiply  (\ref{eq:38-1}) by $\Delta_{\sigma\tau}p_a^\sigma p_a^\tau$ and integrate over $\sqrt{-g}\frac{d^{3}p_{a}}{p_{a0}}$. For this purpose, we use the integrals from the Appendix. We also eliminate the derivative projections $U^\mu\partial_\mu$ with the help of the partial particle number density and energy per particle balance equations. Such balance equations correspond to an Eulerian fluid, where non-equilibrium quantities  $\J_{a}^{\mu}, \q_a^{\mu} \varpi_a, \p_a^{\langle\mu\nu\rangle}$ vanish, that is
\ben\lb{ns0}
U^\mu\partial_\mu \n_a+\n_a\nabla^\mu U_\mu=0,\\ \n_a \c^a_v U^\mu\partial_\mu T+\p_a \nabla^\mu U_\mu=0.
\een
The result of this process becomes a system of equations for $\varpi_b$ and it reads
\ben\lb{ns1}
-\left[\frac{\p_a kT}{c^3}\frac{\partial\ln\z_a}{\partial \ln\c_v^a}\right]\nabla_\mu U^\mu=\sum_{b=1}^r \mathcal{R}_{ab}\varpi_b.
\een
Here we have introduced the matrix $\mathcal{R}_{ab}$, which is defined for different indices $\left\{ a,b\right\}$ as:
\ben\no
\mathcal{R}_{ab}=\frac{U_\mu U_\nu U_\sigma }{c^2\p_b}\int p_a^\mu p_a^\nu \mathcal{I}_{ab}\Bigg[\frac{\partial\ln\z_b}{\partial \ln\c_v^b}\Bigg(\frac{ U_\tau p_b^\tau}{kT}
\\\lb{ns2}
-\frac{3(\c_p^b+\h_b/T)}{\c_v^b}\Bigg)\frac{p_b^\sigma}{kT}
\Bigg]\sqrt{-g}\frac{d^{3}p_{a}}{p_{a0}},\qquad a\neq b.
\een
And in a similar fashion of the previous sections, we write this matrix for equal indices $\left\{ a,b=a\right\}$ as
\ben\no
\mathcal{R}_{aa}=\frac{U_\mu U_\nu U_\sigma }{c^2\p_a}\Bigg\{\sum_{b=1}^r\int p_a^\mu p_a^\nu \mathcal{I}_{ab}\Bigg[\frac{\partial\ln\z_a}{\partial \ln\c_v^a}\Bigg(\frac{ U_\tau p_a^\tau}{kT}
\\\no
-\frac{3(\c_p^a+\h_a/T)}{\c_v^a}\Bigg)\frac{p_a^\sigma}{kT}
\Bigg]
+\int p_a^\mu p_a^\nu \mathcal{I}_{aa}\Bigg[\frac{\partial\ln\z_a}{\partial \ln\c_v^a}\Bigg(\frac{ U_\tau p_a^\tau}{kT}
\\\lb{ns3}
-\frac{3(\c_p^a+\h_a/T)}{\c_v^a}\Bigg)\frac{p_a^\sigma}{kT}
\Bigg]\Bigg\}\sqrt{-g}\frac{d^{3}p_{a}}{p_{a0}}.\qquad
\een

The solution of the linear system of equations (\ref{ns1}) for the partial non-equilibrium pressures $\varpi_a$ is given by
\ben\lb{ns4}
\varpi_a=-\left[\sum_{b=1}^r \left(\mathcal{R}^{-1}\right)_{ab}\frac{\p_b kT}{c^3}\frac{\partial\ln\z_b}{\partial \ln\c_v^b}\right]\nabla_\mu U^\mu,
\een
where  $\left(\mathcal{R}^{-1}\right)_{ab}$ denotes the inverse of the matrix $\mathcal{R}_{ab}$. The constitutive equation for the non-equilibrium pressure of the mixture is obtained from the sum of (\ref{ns4}) over all constituents according to (\ref{sum})$_3$. Hence it follows
\ben\lb{ns5}
\varpi=-\eta\nabla_\mu U^\mu,
\een
where the bulk viscosity  coefficient of the mixture reads
\ben
\eta=\sum_{a,b=1}^r \left(\mathcal{R}^{-1}\right)_{ab}\frac{\p_b kT}{c^3}\frac{\partial\ln\z_b}{\partial \ln\c_v^b}.
\een

For the second equation that conforms the Navier-Stokes law, which is the pressure deviator constitutive one, we proceed in an analogous manner. We take the product of (\ref{eq:38-1}) with $[\Delta_{\sigma}^{(\mu}\Delta_\tau^{\nu)}-\Delta_{\sigma\tau}\Delta^{\mu\nu}/3]p_a^\sigma p_a^\tau$ and integrate over $\sqrt{-g}\frac{d^{3}p_{a}}{p_{a0}}$. This process leads to the following linear system of equations for the  partial pressure deviator tensors $\p^{\langle\mu\nu\rangle}_b$:
\ben\lb{ns6}
2\nabla^{\langle\mu}U^{\nu\rangle} =\sum_{b=1}^r\mathcal{K}_{ab}\p_b^{\langle\mu\nu\rangle}.
\een
In this last equation we have introduced the following abbreviation for the symmetric and traceless four-velocity gradient
\ben
\nabla^{\langle\mu}U^{\nu\rangle}=\left( \frac{\Delta^\mu_\sigma\Delta^\nu_\tau+\Delta^\nu_\sigma\Delta^\mu_\tau}2-\frac{\Delta^{\mu\nu}\Delta_{\sigma\tau}}3\right)\partial^\sigma U^\tau.
\een
Equation (\ref{ns6}) also includes the definition of the matrix  $\mathcal{K}_{ab}$ which reads
\ben\no
\mathcal{K}_{ab}=-\frac{c^3\Delta_{\mu\langle\sigma}\Delta_{\tau\rangle\nu}}{10\p_a \h_a\p_b}
\int p_a^\sigma p_a^\tau\mathcal{I}_{ab}\left[\frac{\z_b}{m_b\h_b}p_b^\mu p_b^\nu\right]
\\\lb{ns7}
\times\sqrt{-g}\frac{d^{3}p_{a}}{p_{a0}},\qquad a\neq b\qquad
\\\no
\mathcal{K}_{aa}=-\frac{c^3\Delta_{\mu\langle\sigma}\Delta_{\tau\rangle\nu}}{10\p_a \h_a\p_a}\Bigg\{
\sum_{b=1}^r\int p_a^\sigma p_a^\tau\mathcal{I}_{ab}\left[\frac{\z_a}{m_a\h_a}p_a^\mu p_a^\nu\right]
\\\lb{ns8}
+\int p_a^\sigma p_a^\tau\mathcal{I}_{aa}\left[\frac{\z_a}{m_a\h_a}p_a^\mu p_a^\nu\right]\Bigg\}\sqrt{-g}\frac{d^{3}p_{a}}{p_{a0}}.\qquad
\een

From the solution of the linear system of equations (\ref{ns6}) for $\p_b^{\langle\mu\nu\rangle}$ and from the relationship $\p^{\langle\mu\nu\rangle}=\sum_{b=1}^r\p_b^{\langle\mu\nu\rangle},$ it follows the constitutive equation for the pressure deviator tensor of the mixture:
\ben\lb{ns9}
\p^{\langle\mu\nu\rangle}=2\mu \nabla^{\langle\mu}U^{\nu\rangle}.
\een
Here the  shear viscosity coefficient of the mixture is given by
\ben
\mu=\sum_{a,b=1}^r \left(\mathcal{K}^{-1}\right)_{ab}.
\een

Equations (\ref{ns5}) and (\ref{ns9}) are the constitutive equations of a relativistic Newtonian fluid, also known as the Navier-Stokes constitutive equations.

\section{Conclusions}

In this work we have studied a mixture of $r$ species of relativistic gases in the presence of gravitational fields. The curvature of the space-time was introduced by incorporating the Christoffel symbols to the Boltzmann equation. We used the Schwarzschild metric written in isotropic coordinates. A linearized Boltzmann equation was obtained by following a methodology
which combines the features of the Chapman-Enskog and Grad methods.

By applying the Chapman-Enskog-Grad combined method to the Boltzmann equation we obtained a linear expression (Eq. (\ref{eq:38-1})) which was used for the determination of the thermodynamic fluxes as functions of the thermodynamic forces. The Navier-Stokes law was derived as well as the generalized of Fourier and Fick laws.

The proof of the validity
of the Onsager reciprocity relations was possible lastly because of the
symmetries of the collisional term of the Boltzmann equation. These
symmetries are those associated with the H-theorem and the microscopic
reversibility principle. This reinforces the idea that the Onsager
reciprocity relations are the macroscopic manifestation of the microscopic
symmetries of the trajectories of the particles that conform the gas.

We have introduced the thermal force
\begin{equation}
\nabla^\mu\mathcal{T}=\nabla^\mu T-\frac{T}{c^{2}}\Delta^{\mu i}\left[U^{\nu}\frac{\partial U_{i}}{\partial x^{\nu}}-\frac{1}{1-{\Phi^2}/4c^{4}}\frac{\partial{\Phi}}{\partial x^{i}}\right],\label{eq:68}
\end{equation}
 which is a very eloquent result. Indeed, Eq. (\ref{eq:68}) turns
to be just the gradient of the temperature in the non-relativistic
limit i.e. $\nabla^\mu T$ because the factor $T/c^{2}$ of the second
term is of relativistic order. The inclusion of the acceleration term into the thermal force was proposed by Eckart \cite{Eck} while the one relating the gravitational potential gradient by Tolman \cite{To1,To2}. Here these terms appear as a natural consequence of the solution of the relativistic Boltzmann equation in gravitational fields.

On the other hand we have identified the generalized diffusion force with
\ben\no
&&\d_a^{\mu}=\nabla^\mu\x_{a}+\left(\x_{a}-1\right)\nabla^\mu\ln\p
\\\lb{dd}
&&-\frac{\n_{a}\h_{a}-\n\h}{\p c^{2}}\Delta^{\mu j}\left[U^{\tau}\frac{\partial U_{j}}{\partial x^{\tau}}-\frac{1}{1-{\Phi^2}/4c^{4}}\frac{\partial{\Phi}}{\partial x^{j}}\right].
\een
 This is a new and interesting result because the third term -- which is related with the  four-acceleration and the gradient of the gravitational potential -- does not go to zero in
the non-relativistic limiting case as the thermal force. As it was pointed out in the work \cite{KD}, the  diffusion force that came out from a non-relativistic kinetic theory \cite{CC,HCB} has a  similar expression  to (\ref{dd}). It depends on the concentration and pressure gradients, but it has a term depending on the forces which act on the particle of different species, such a term vanishes when only gravitational forces are acting on the particles.

Another result obtained is the entropy flux of the relativistic mixture through the use of Grad's distribution function, which has a similar  expression as the one of non-relativistic Linear Irreversible Thermodynamics \cite{GM}. Its constitutive equation was written in terms of the generalized thermal and diffusion forces, so that it depends also on the acceleration and on the gravitational potential gradient.

Here is the place to discuss two  additional issues. The first one is the validity of the Onsager reciprocity relations for the case of a relativistic quantum gas. In such a case, the local equilibrium distribution (which in this work is given by Eq. (\ref{eq:7})) would take a form of the Fermi-Dirac and Bose-Einstein distributions for a fermionic and bosonic gas, respectively. Quantum relativistic gases can be described by the relativistic Uehling-Uhlenbeck equation (see e. g. \cite{CK}). As we have pointed out, the validity of Onsager's reciprocity relations are deeply associated with the symmetries that belong to the H-theorem. In the present work, those symmetries are implied in Eq. (\ref{eq:32}). Then, to show the validity of the Onsager reciprocity relations for a quantum system we need the validity of the H-theorem, fact that has been shown in the literature in Refs. \cite{CerKre} and \cite{CKHT} raising the possibility to explore that issue. The second topic is related with Tolman's law, which has been derived in \cite{To1} and is valid for all  static spherical symmetrical line element. In the present work we have used the Schwarzschild metric, which according to Birkoff's theorem is the  most general spherically symmetrical non-rotating and uncharged source of the gravitational field.

As a final comment we call attention to the fact that for the determination of all the transport coefficients, we have to  specify the interaction potential of the relativistic particles and  evaluate  the matrices $\left\{ \mathcal{R}_{ab}, \mathcal{K}_{ab},\mathcal{A}_{ab},
\mathcal{F}_{ab}, \mathcal{H}_{ab}\right\} $. This represents work in progress and will be published in the future.

\section*{Acknowledgment}

V. M. acknowledges the CONACyT-M\'exico and SECITI-CLAF M\'exico-Brazil for financial support
 and G. M. K. the Conselho Nacional de Desenvolvimento Cient\'{\i}fico e Tecnol\'ogico (CNPq), Brazil.

\section*{Appendices}
\subsection*{ Table of integrals}

For the purposes of this work, it is convenient to do the following
unique decomposition of the integral operators:
\begin{widetext}
\ben
X_{ab}^{\mu\nu}=\int p_a^\mu \mathcal{I}_{ab}\left[p_b^\nu\right]\sqrt{-g}\frac{d^{3}p_{a}}{p_{a0}}=
\frac1{3c^2}\left(4I_{ab}^1-I_{ab}^2\right)U^\mu U^\nu +\frac1{3}\left(I_{ab}^2-I_{ab}^1\right)g^{\mu\nu},
\\
X_{ab}^{\mu\nu\sigma}=\int p_a^\mu p_a^\nu\mathcal{I}_{ab}\left[p_b^\sigma\right]\sqrt{-g}\frac{d^{3}p_{a}}{p_{a0}}=
\frac2{3c^3}\left(3I_{ab}^3-I_{ab}^4\right)U^\mu U^\nu U^\sigma -\frac1{3c}I_{ab}^3g^{\mu\nu}U^\sigma\cr+\frac1{3c}\left(I_{ab}^4-I_{ab}^3\right)\left(g^{\mu\sigma}U^\nu+g^{\nu\sigma}U^\mu\right),
\\
X_{ab}^{\mu\nu\sigma\tau}=\int p_a^\mu p_a^\nu\mathcal{I}_{ab}\left[p_b^\sigma p_b^\tau\right]\sqrt{-g}\frac{d^{3}p_{a}}{p_{a0}}=
\frac2{15c^4}\left(24I_{ab}^5-12I_{ab}^6+I_{ab}^7\right)U^\mu U^\nu U^\sigma U^\tau  +\frac1{15c^2}(I_{ab}^7-2I_{ab}^6
\cr
-6I_{ab}^5)(g^{\mu\nu}U^\sigma U^\tau+g^{\sigma\tau} U^\mu U^\nu)+\frac1{30c^2}\left(16I_{ab}^6-3I_{ab}^7-12I_{ab}^5\right)(g^{\mu\sigma}U^\nu U^\tau+g^{\mu\tau}U^\nu U^\sigma+g^{\nu\sigma}U^\mu U^\tau
\cr
+g^{\nu\tau}U^\mu U^\sigma)+\frac1{30}\left(3I_{ab}^7-6I_{ab}^6+2I_{ab}^5\right)\left(g^{\mu\sigma}g^{\nu\tau}+g^{\mu\tau}g^{\nu\sigma}\right)+\frac1{15}\left(I_{ab}^5-I_{ab}^7+2I_{ab}^6\right)g^{\mu\nu}g^{\sigma\tau},
\een
where $I_{ab}^1\dots I_{ab}^7$ are given by
\ben
I_{ab}^1=\frac{U_\mu U_\nu}{c^2} X_{ab}^{\mu\nu},\qquad I_{ab}^2=g_{\mu\nu} X_{ab}^{\mu\nu},\qquad
I_{ab}^3=\frac{U_\mu U_\nu U_\sigma}{c^3} X_{ab}^{\mu\nu\sigma},\qquad I_{ab}^4=\frac{g_{\mu\sigma} U_\nu}{c}
 X_{ab}^{\mu\nu\sigma},\\
 I_{ab}^5=\frac{U_\mu U_\nu U_\sigma U_\tau}{c^4} X_{ab}^{\mu\nu\sigma\tau},\qquad I_{ab}^6=\frac{g_{\nu\tau} U_\mu U_\sigma}{c^2} X_{ab}^{\mu\nu\sigma\tau},\qquad I_{ab}^7=g_{\mu \sigma}g_{\nu\tau}  X_{ab}^{\mu\nu\sigma\tau}.
\een
\end{widetext}

Here  we list a table of integrals that are used in the previous sections.
\begin{widetext}
\ben
&&\int e^{-{1\over kT}U^\lambda p_\lambda}{d^3p\over p_0}=4\pi m kT
K_1(\zeta),\qquad
\int p^{\mu}e^{-{1\over kT}U_\lambda p^\lambda}{d^3p\over p_0}=
4\pi m^2k T K_2(\zeta)U^{\mu},
\\
&&\int p^{\mu}p^{\nu}e^{-{1\over kT}U_\lambda p^\lambda}
{d^3p\over p_0}=-4\pi (mk T )^2\Big[K_2(\zeta)g^{\mu\nu}
-\z K_3(\zeta)\frac{U^{\mu}U^{\nu}}{c^2}\Big],
\\
&&\int p^{\mu}p^{\nu}p^{\sigma}
e^{-{1\over kT}U_\lambda p^\lambda}{d^3p\over p_0}=
-4\pi m^3 (k T )^2\Big[\frac{K_3(\zeta)}3g^{(\mu\nu}U^{\sigma)}
-\z K_4(\zeta)\frac{U^{\mu}U^{\nu}U^{\sigma}}{c^2}\Big],
\\
&&\int p^{\mu}p^{\nu}p^{\sigma}p^{\tau}
e^{-{1\over kT}U_\lambda p^\lambda}{d^3p\over p_0}=
4\pi (mk T )^3\bigg[\frac{K_3(\zeta)}3
g^{(\mu\nu}g^{\sigma\tau)}-\z K_4(\z)\frac{g^{(\mu\nu}U^{\sigma}U^{\tau)}}{6c^2}
+\z^2K_5(\zeta)\frac{U^{\mu}U^{\nu}U^{\sigma}U^{\tau}}{c^4}
\bigg],
\\\no
&&\int p^{\mu}p^{\nu}p^{\sigma}
p^{\tau}p^{\epsilon}
e^{-{1\over kT}U_\lambda p^\lambda}{d^3p\over p_0}=
4\pi m^4(k T )^3\bigg[\frac{K_4(\zeta)}{15}U^{(\epsilon}g^{\mu\nu}
g^{\sigma\tau)}
-\z K_5(\zeta)
\frac{g^{(\mu\nu}U^{\sigma}U^{\tau}U^{\epsilon)}}{10c^2}
\\
&&+\z^2K_6(\zeta)\frac{U^{\mu}U^{\nu}U^{\sigma}U^{\tau}
U^{\epsilon}}{c^4}\bigg],
\\
&&\int \frac{e^{-{1\over kT}U^\lambda p_\lambda}}{U^\tau p_\tau}{d^3p\over p_0}=4\pi m\left[K_1(\zeta)-{\rm Ki}_1(\zeta)\right],\qquad \int p^\mu\frac{e^{-{1\over kT}U^\lambda p_\lambda}}{U^\tau p_\tau}{d^3p\over p_0}=4\pi m^2\frac{K_1(\zeta)}{\z}U^\mu,
\\\no
&&\int p^\mu p^\nu\frac{e^{-{1\over kT}U^\lambda p_\lambda}}{U^\tau p_\tau}{d^3p\over p_0}=-\frac{4\pi m^2kT}{3}\bigg\{\big[K_2(\z)-\z\big(K_1(\zeta)-{\rm Ki}_1(\zeta)\big)\big]g^{\mu\nu}
\\
&&-\frac{1}{c^2}\big[4K_2(\z)-\z\big(K_1(\zeta)-{\rm Ki}_1(\zeta)\big)\big]U^\mu U^\nu\bigg\},
\\
&&\int p^\mu p^\nu p^\sigma\frac{e^{-{1\over kT}U^\lambda p_\lambda}}{U^\tau p_\tau}{d^3p\over p_0}=-\frac{4\pi m^2k^2T^2}{c^2}\bigg\{
\frac{K_2(\z)}3g^{(\mu\nu}U^{\sigma)}
-\big[\z K_3(\z)+2K_2(\z)\big]\frac{U^\mu U^\nu U^\sigma}{c^2}\bigg\},
\\\no
&&\int p^\mu p^\nu p^\sigma p^\tau\frac{e^{-{1\over kT}U^\lambda p_\lambda}}{U^\theta p_\theta}{d^3p\over p_0}=\frac{4\pi m^3k^2T^2}{15}\bigg\{
\frac{3K_3(\z)-\z K_2(\z)+\z^2[K_1(\zeta)-{\rm Ki}_1(\zeta)]}{3}g^{(\mu\nu}g^{\sigma\tau)}
\\\no
&&-\frac{1}{6c^2}\left[18K_3(\z)-\z K_2(\z)+\z^2\left(K_1(\zeta)-{\rm Ki}_1(\zeta)\right)\right]g^{(\mu\nu}U^\sigma
U^{\tau)}
\\
&&+\frac{3}{c^4}\bigg[48K_3(\z)+4\z K_2(\z)+\z^2\left(K_1(\zeta)-{\rm Ki}_1(\zeta)\right)\bigg]U^\mu U^\nu U^\sigma U^\tau\bigg\},
\\\no
&&\int p^\mu p^\nu p^\sigma p^\tau p^\epsilon\frac{e^{-{1\over kT}U^\lambda p_\lambda}}{U^\theta p_\theta}{d^3p\over p_0}=\frac{4\pi m^6c^4}{\z^3}\bigg\{
\frac{K_3(\z)}{15}U^{(\mu}g^{\nu\tau}g^{\sigma\epsilon)}-\frac{1}{10c^2}[8K_3(\z)+\z K_2(\z)]
g^{(\mu\epsilon}U^\nu U^\sigma U^{\tau)}
\\
&&
+\frac{1}{c^4}\big[\z^2K_3(\z)+12\z K_2(\z)+80K_3(\z)\big]
U^\mu U^\nu U^\sigma U^\tau U^\epsilon\bigg\}.
\een
Above the parenthesis around $N$ indexes indicate a sum over all permutations of these indexes divided by $N!$. Furthermore, ${\rm Ki}_n(\zeta)$ denotes the integral
\ben
{\rm Ki}_n(\zeta)=
\int_0^\infty {e^{-\zeta\cosh t}\over \cosh^nt}dt.
\een
\end{widetext}
\subsection*{Cristoffel symbols for the Schwarzschild isotropic metric}
\begin{widetext}
 \ben\lb{a4a}
&& \Gamma_{00}^0=0,\qquad \Gamma_{ij}^0=0,\qquad \Gamma_{ij}^k=0 \quad (i\neq j\neq k),\qquad
\Gamma_{0j}^i=0,\qquad \Gamma_{\underline{i}\,j}^{\underline{i}}=\frac{1}{2g_1(r)}\frac{d g_1(r)}{dr}\delta_{jk} \frac{x^k}{r},
 \\\lb{a4b}
  &&
 \qquad\Gamma_{0i}^0=\frac{1}{2g_0(r)}\frac{d g_0(r)}{dr}\delta_{ij}
 \frac{x^j}{r},
 \qquad\Gamma_{00}^i=\frac{1}{2g_1(r)}\frac{d g_0(r)}{dr}\frac{x^i}{r},\qquad
 \Gamma_{\underline{i}\,\underline{i}}^j=-\frac{1}{2g_1(r)}\frac{d g_1(r)}{dr}\frac{x^j}{r}\quad (i\neq j).
 \een
The underlined indices above are not summed and
\ben\lb{a5a}
\frac{d g_0(r)}{dr}=\frac{2GM}{c^2r^2}\frac{\left(1-\frac{GM}{2c^2r}\right)}{\left(1+\frac{GM}{2c^2r}\right)^3},\qquad
\frac{d g_1(r)}{dr}=-\frac{2GM}{c^2r^2}\left(1+\frac{GM}{2c^2r}\right)^3.
\een
\end{widetext}

\end{document}